\documentstyle[12pt]{article}

\newcommand{\be}{\begin{equation}}
\newcommand{\ee}{\end{equation}}

\def\psnormal{\textwidth=16cm\textheight=21.5cm
          \oddsidemargin=0.5cm\evensidemargin=0cm
          \topmargin=0cm\parindent=1cm}
\psnormal

\begin{document}
\pagestyle{empty}

\hspace{3cm}

\vspace{-3.4cm}
\rightline{{ CERN--TH.6681/92}}
\rightline{{ IEM--FT--61/92}}
\rightline{{ FTUAM 92/31}}

\vspace{0.4cm}
\begin{center}
{\bf {\large S}OFT {\large SUSY} {\large B}REAKING {\large T}ERMS
IN {\large S}TRINGY {\large S}CENARIOS: {\large C}OMPUTATION
AND {\large P}HENOMENOLOGICAL {\large V}IABILITY}
\vspace{0.8cm}

B. de CARLOS${}^*$, J.A. CASAS${}^{**,*}\;$ and C. MU\~NOZ${}^{***}$
\vspace{0.8cm}

${}^{*}$ Instituto de Estructura de la Materia (CSIC),\\
Serrano 123, 28006--Madrid, Spain
\vspace{0.4cm}

${}^{**}$ CERN, CH--1211 Geneva 23, Switzerland
\vspace{0.4cm}

${}^{***}$ Dept. de F!sica Te\'orica C-XI, \\
Univ. Aut\'onoma de Madrid, E-28049 Madrid, Spain
\vspace{0.4cm}

\end{center}

\centerline{\bf Abstract}
\vspace{0.3cm}

\noindent
We calculate the soft SUSY breaking terms arising from a
large class of string scenarios, namely symmetric orbifold constructions,
and study its phenomenological
viability. They exhibit a certain lack of universality, unlike
the usual assumptions of the minimal supersymmetric standard model.
Assuming gaugino condensation in the hidden sector as the source
of SUSY breaking, it turns out that squark and slepton masses tend
to be much larger than gaugino masses. Furthermore, we show
that these soft breaking terms can be  perfectly consistent
with both experimental and naturalness constraints
(the latter comes from the absence of fine tuning
in the $SU(2)\times U(1)_Y\rightarrow U(1)_{em}$ breaking process).
This is certainly non--trivial and in fact imposes interesting constraints
on measurable quantities. More precisely,  we find that the gluino mass
($M_3$) and the chargino mass ($M_{\chi^{\pm}}$) cannot be much higher
than their present experimental lower bounds
($M_3\stackrel{<}{{}_\sim}285\ $GeV ;
$M_{\chi^\pm}\stackrel{<}{{}_\sim}80\ $GeV),
while squark and slepton masses must be much larger
($\stackrel{>}{{}_\sim} 1\ $TeV). This can be
considered as an observational signature of this kind of
stringy scenarios. Besides, the top mass is constrained to be
within a range
($80\ $GeV$\stackrel{<}{{}_\sim}m_t\stackrel{<}{{}_\sim}165\ $GeV)
remarkably consistent with its present experimental bounds.

\vspace{0.3cm}
\begin{flushleft}
{CERN--TH.6681/92} \\
{IEM--FT--61/92} \\
{FTUAM 92/31} \\
{October 1992}
\end{flushleft}
\psnormal
%
%

\newpage
\pagestyle{plain}
\pagenumbering{arabic}
\section{Introduction}

In the last two years there has been substantial progress in our
understanding of supersymmetry (SUSY) breaking in string theories
[1--11]. In particular, it has been learned that gaugino condensation
effects in the hidden sector are able to break SUSY at a hierarchically
small scale, at the same time as the dilaton $S$ and the moduli $T_i$
acquire reasonable vacuum expectation values (VEVs).
(The VEV of $S$ is related to the value of the gauge coupling constant,
while those of $T_i$ define the size and shape of the compactified space.)
This has been realized in the context of symmetric orbifold constructions.
The modular invariance in the target space [3--6] as well as the presence
of matter in the hidden sector play an important role in this process
\cite[7--11]{cas1}. On the other hand, the soft SUSY breaking terms in
the observable sector are the phenomenological signature of any SUSY
breaking mechanism.
Therefore, it is essential to completely calculate the soft breaking terms
of these scenarios and study their phenomenological viability.
This is the motivation of the present paper.

There are two types of tests that the soft breaking terms should pass.
First, they have to be consistent with the experimental (lower) bounds
on gaugino masses, squark masses, etc. Second, they should be small enough
not to spoil the SUSY solution to the gauge hierarchy problem, guaranteeing
a successful $SU(2)\times U(1)_Y\rightarrow U(1)_{em}$ breaking. The latter
is a naturalness requirement. We find in this paper that SUSY breaking
by gaugino condensation in the string models considered
can be perfectly consistent with both tests. Moreover,
this imposes very interesting constraints on the parameters defining
the SUSY breakdown. As a consequence, we find that the gluino mass
($M_3$) and the lightest chargino mass ($M_{\chi^{\pm}}$) cannot be much higher
than their present experimental lower bounds
($M_3\stackrel{<}{{}_\sim}285\ $GeV,
$M_{\chi^\pm}\stackrel{<}{{}_\sim}80\ $GeV),
while squark and slepton masses must be much larger
($\stackrel{>}{{}_\sim} 1\ $TeV). This can be
considered as an experimental signature of this kind of models.
Also, the top mass is constrained to be
within a range ($80\ $GeV$\stackrel{<}{{}_\sim}m_t\stackrel{<}{{}_\sim}
165\ $GeV) that is
remarkably consistent with its present experimental bounds.
In the calculation we have assumed minimal particle content in the
observable sector and the vanishing of the cosmological constant.
In section 2 we give the general form of the soft SUSY breaking terms
for symmetric orbifold constructions. Interestingly enough, they
show a certain lack of universality, unlike the usual
assumptions of the minimal supersymmetric standard model.
In order to get more quantitative results a specific SUSY breaking
mechanism has to be considered. This is done in section 3
assuming the above mentioned gaugino condensation in the hidden sector
(the only mechanism so far analized capable to generate
a hierarchical SUSY breakdown in string constructions).
In section 4 we confront the corresponding soft breaking terms with
the two types of tests mentioned above. As a consequence we obtain
the allowed ranges of variation for different physical
quantities. Finally, in section 5 we present our conclusions.

\section{General characteristics}

Following the standard notation we define the soft breaking terms
in the observable sector by
\begin{eqnarray}
{\cal L} = {\cal L}_{{\mathrm {SUSY}}}+{\cal L}_{{\mathrm {soft}}}\;\;.
\label{Lobs}
\end{eqnarray}
Here ${\cal L}_{{\mathrm {SUSY}}}$ is the supersymmetric Lagrangian derived
from the observable superpotential $W_{{\mathrm {obs}}}$, which includes
the usual Yukawa terms $W_Y$ and a mass coupling $\mu H_1H_2$
between the two Higgs doublets $H_1$, $H_2$. Assuming canonically
normalized fields, ${\cal L}_{{\mathrm soft}}$ is given by
\begin{eqnarray}
{\cal L}_{{\mathrm{soft}}} = -\sum_\alpha m_{\alpha}^2 |\phi_\alpha|^2
-\frac{1}{2} \sum_{a=1}^3 M_a \bar\lambda_a\lambda_a
-\left(Am_{3/2}W_Y + Bm_{3/2}\mu H_1H_2
 \ +\ \mathrm{h.c.}\right)
\label{Lsoft}
\end{eqnarray}
where $m_{3/2}$ is the gravitino mass, $\phi_\alpha$
represent the scalar components of the
supersymmetric particles, and $\lambda_a$ are the $U(1)_Y,
SU(2),SU(3)$ gauginos.

The characteristics of soft SUSY breaking terms in the observable
sector are, to a great extent, determined by the type of $N=1$ SUGRA
theory which appears in four dimensions. As it is known, this is
characterized by the gauge kinetic function $f$, the K\"{a}hler
potential $K$ and the superpotential $W$. It is also customary
to define ${\cal G}=K+\log|W|^2$. These
functions are determined, in principle, in a given
compactification scheme, although in practice they are sufficiently
well known only for orbifold compactification schemes \cite{dix2}, which
on the other hand have proved to possess very attractive features
from the phenomenological point of view \cite{cas2}. More precisely,
in the general case when the gauge group contains several factors $G =
\prod_{a} G_{a}$, the exact gauge kinetic functions in string
perturbation theory, up to small field--independent contributions,
are [14--16]
\begin{eqnarray}
f^a = k^aS + \frac{1}{4\pi^2}\sum_{i=1}^3 (\frac{1}{2}b_i^{'a} - k^a
\delta_{i}^{GS}) \log(\eta(T_{i}))^{2}\;\;,
\label{fw}
\end{eqnarray}
with
\begin{eqnarray}
b_{i}^{'a} = C(G^a)-\sum_\Phi T(R_\Phi^a)(1+2n_\Phi^i)\;\;,
\label{bp}
\end{eqnarray}
where the meaning of the various quantities appearing in
eqs.(\ref{fw},\ref{bp}) is the following:
$k^{a}$ is the Kac--Moody level of the $G^{a}$ group ($k^a=1$ is
a very common possibility), $S$ is the dilaton field,
$T_{i}$ ($i = 1, 2, 3$) are untwisted moduli whose real parts
give the radii of the three compact complex dimensions of the
orbifold (Re $T_{i}\propto R_{i}^{2}, i = 1, 2, 3$),
$\delta_{i}^{GS}$ are 1--loop contributions coming from the
Green--Schwarz mechanism, which have been
determined for the simplest (2,2) $Z_{N}$ orbifolds \cite{der};
$\eta(T_{i})$ is the Dedekind function, $\Phi$ labels the matter fields
transforming as $R^a_\Phi$ representations under $G^a$, and $n_\Phi^i$ are the
corresponding modular weights. $C(G)$ denotes the Casimir operator
in the adjoint representation of $G$ and $T(R)$ is defined by
Tr$(T^iT^j)=T(R)\delta^{ij}$.
The K\"{a}hler potential $K$ is given by \cite{dix3,der}
\begin{eqnarray}
K = K_o(S,T_i)\ +\ \sum_\alpha K_1^\alpha |\phi_\alpha|^2\
+\ O(|\phi_\alpha|^4)
\;\;,
\label{K}
\end{eqnarray}
where $\phi_\alpha$ are the matter fields and
\begin{eqnarray}
K_o(S,T_i) & =&  -\log( S + \bar S + \frac{1}{4\pi^2}\sum_{i=1}^3
\delta_{i}^{GS}
\log(T_{i} + \bar{T_{i}})) - \sum_{i=1}^3 \log(T_{i} + \bar{T_{i}})
\nonumber \\
& \equiv & -\log Y - \sum_{i=1}^3 \log(T_{i} + \bar{T_{i}})
\nonumber \\
K_1^\alpha &=& \prod_i (T_i+\bar T_i)^{n_\alpha^i}
\;\;.
\label{K01}
\end{eqnarray}
The function $Y$ can be considered \cite{der}
as the redefined gauge coupling constant (up to threshold
corrections) at the unifying string scale
($Y=2g_{\mathrm{str}}^{-2}$). $K_o$ is known at the 1--loop
order, while $K_1^{\alpha}$ is only known at tree level. The complete
dependence of $K$ (at tree level) on untwisted matter fields is
also known \cite{wit}, while for twisted fields this dependence has been
conjectured in ref.\cite{fer2}. However, for our purposes, i.e. studying
soft breaking terms, the above expressions are sufficient. The
1--loop dependence on untwisted fields has also been conjectured
in ref.\cite{lu2}
\begin{eqnarray}
K_{\mathrm{1-loop}}^{\mathrm{untw}} = -\log \left[ S + \bar{S}
+ \frac{1}{4\pi^2}\sum_{i}
\delta_i^{GS}\log X_i \right] - \sum_{i=1}^3 \log X_i
\;\;,
\label{Kuntw}
\end{eqnarray}
where $X_i=(T_i + \bar{T_i}-\sum_x |A_i^x|^2)$, with $A_i^x$ denoting
the untwisted fields (recall that untwisted fields carry a holomorphic
index $i$). We will see below that (\ref{K}) and (\ref{Kuntw}) yield
practically equivalent results.
Finally, the perturbative superpotential $W^{p}$ at the
renormalizable level has the form
\begin{eqnarray}
W^{p} = h_{IJK} \Phi^{(u)}_{I}\Phi^{(u)}_{J}\Phi^{(u)}_{K} +
h^{'}_{IJK}(T_{i}) \Phi^{(t)}_{I}\Phi^{(t)}_{J}\Phi^{(t)}_{K} +
h^{''}_{IJK} \Phi^{(u)}_{I}\Phi^{(t)}_{J}\Phi^{(t)}_{K}
\;\;,
\label{Wpert}
\end{eqnarray}
where $\Phi^{(u)}_{I}$ $(\Phi^{(t)}_{I})$ are untwisted (twisted) charged
matter fields.
The value of $h_{IJK}$, $h^{''}_{IJK}$ for the allowed couplings
is simply a constant, while $h^{'}_{IJK}(T_{i})$ are complicated but
known functions of $T_{i}$ \cite{ham}. Besides $W^{p}$, there is a
non--perturbative piece, $W^{np}$, usually triggered by gaugino
condensation effects in the hidden sector, which is crucial to break
SUSY. In the following
we will assume $\langle W^{p}\rangle =0$, $\langle W^{np}\rangle\neq 0$,
as it happens in all
SUSY breaking scenarios so far analysed. $W^{np}$ depends on the
$S$ and $T$ fields and, sometimes, on certain matter fields $A$, which
are singlet under the relevant gauge groups [7--11]. As has been
shown  \cite{priv,dec2},
the condition $\partial W/\partial A=0$
is the correct one to integrate out these fields. In consequence, we can
use
\begin{eqnarray}
W^{np} = W^{np}(S,T_i)
\label{Wnp}
\end{eqnarray}
without any loss of generality. Expressions (\ref{fw}--\ref{Wnp}) are to
be understood at the string scale $M_{Str}=0.527\times g \times 10^{18}
\ $GeV \cite{kapl}, where $g\simeq 1/\sqrt{2}$ is the corresponding value of
the gauge coupling constant.

The next step is to calculate, once SUSY breaking is assumed, the
form of the soft breaking terms based on eqs.(\ref{fw}--\ref{Wnp}).
Some work in this direction has already been done. The gaugino masses
were evaluated in ref.\cite{font} for $\delta_i^{GS}=0$, and
in ref.\cite{ib1} for $\delta_i^{GS}\neq 0$ and a generic form of
the  K\"{a}hler potential.
Likewise, the scalar masses when
$\delta_i^{GS}=0$ were calculated in ref.\cite{ib1}, while the form of
trilinear terms for $\delta_i^{GS}=0$
was given in ref.\cite{cve}. It is interesting to note that
$\delta_i^{GS}\neq 0$ turns out to be crucial for a viable
phenomenology, as will be seen in the next sections. In this paper
we have calculated the general expressions for all the soft breaking terms.
The notation used is that of ref.\cite{crem}. We work with the usual
overall modulus simplification $T=T_1=T_2=T_3$.

\vspace{0.3cm}
\noindent {\em Gaugino masses:}

Physical gaugino masses $M_a$ for the canonically normalized
gaugino fields ($\lambda_{phys}=
(\mathrm{Re}\ f)^{1/2}\lambda$) are given by
$\frac{1}{2}(\mathrm {Re}\ f)^{-1}\sum_\alpha f_a^\alpha e^{{\cal G}/2}
({\cal G}^{-1})^{\bar \beta}_\alpha {\cal G}_{\bar \beta}$.
Using the renormalization
group (RG) equation for $M_a$, i.e. $M_a(Q)=
\frac{\alpha_a(Q)}{\alpha_a(M_{Str})}M_a(M_{Str})$,
and $\mathrm{Re}\ f_a=g_a^{-2}(M_{Str})=[4\pi\alpha_a(M_{Str})]^{-1}$,
we directly write $M_a$ at any $Q$ scale
\begin{eqnarray}
M_a(Q) & = & 2\pi\alpha_{a}(Q) m_{3/2} \left\{ k^{a}Y^{2}
\left(-\frac{1}{Y}+\frac{\bar{W}_{\bar S}}{\bar W}\right)
+ \frac{Y(T+\bar T)^2}{3Y+\frac{\delta^{GS}}{4 \pi^2}}
\frac{\hat{G}_{2}(T)}{\pi}
\right.
 \nonumber \\
 & \times  & \left.
\left(\sum_{i} \frac{b_i^{'a}}{16 \pi^2}-
k^a\frac{\delta^{GS}}{8 \pi^2}\right)
\left[ \frac{1}{T+\bar T}
\left(3+\frac{\delta^{GS}}{4 \pi^2}
\frac{\bar{W}_{\bar S}}{\bar W}\right)
-\frac{\bar W_{\bar T}}{\bar W}\right]\right\}
\;\;,
\label{M12}
\end{eqnarray}
%
where $W_\phi=\partial W/\partial\phi$, $\phi=S,T$,
$m_{3/2}=e^{K/2}|W|$, $\delta^{GS}=\sum_i \delta^{GS}_i$
and $\hat G_2(T)=-(\frac{2\pi}{T+\bar T}+4\pi\eta^{-1}
\frac{\partial\eta}{\partial T})$.
Furthermore it is
worth mentioning that the factors $\sum_i b_{i}^{'a}\hat G_2(T)$
in (\ref{M12}) come from replacing $\sum_i b_{i}^{'a} G_2(T)$ factors (with
$G_2(T)=\hat G_2(T)+\frac{2\pi}{T+\bar T}$) in the original
expression. This
counts the radiative contribution of the massless fields of the theory,
and has to be performed in order to maintain the modular invariance
of $M_a$ \cite{ib1}.

\vspace{0.3cm}
\noindent {\em Scalar masses:}

Scalar masses $m_{\tilde{\phi}_\alpha}$ are obtained from the scalar potential
$V=e^{{\cal G}}({\cal G}_\alpha ({\cal G}^{-1})^{\bar \alpha}_\beta
{\cal G}^{\bar \beta}-3)$.
%
%
They have to be normalized by taking into
account that the $\phi_\alpha$ kinetic terms are $(K_{\alpha \bar \beta})
D\phi_\alpha D\bar \phi_\beta$. Thus we obtain
\begin{eqnarray}
m^2_{{\phi}_{\alpha}} = V_o\ +\ m_{3/2}^{2}\left[
1+\frac{n_{\alpha}Y^{2}}{(3Y+\frac{\delta^{GS}}{4\pi^2})^{2}}
\left| 3+\frac{\delta^{GS}}{4\pi^2}\frac{W_S}{W}-(T+\bar T)\frac{W_T}{W}
\right|^2 \right]
%
%
\;\;,
\label{mQ}
\end{eqnarray}
where $V_o=\langle V\rangle$ is simply the value
of the cosmological constant, which we will assume to vanish through
the paper. One--loop corrections to the previous expressions are, in all
probability, quite small\footnote{For instance, scalar masses for untwisted
fields computed from the conjecture of eq.(\ref{Kuntw}) are given by
$m_\phi^2= V_o\ +\ m_{3/2}^{2}\left[
1-\frac{Y}{(3Y+a)^2}(Y+\frac{a^2}{3Y+a})
\left|3+a\frac{W_S}{W}-(T+\bar T)\frac{W_T}{W} \right|^2
-\frac{1}{3Y+a}(1+\frac{3Y}{3Y+a})a|Y\frac{W_S}{W}-1|^2
\right]$, with $a=\frac{\delta^{GS}}{4\pi^2}$. Numerically,
this amounts to $1\%$ departure from the result of eq.(\ref{mQ}).}.

\vspace{0.3cm}
\noindent {\em Trilinear scalar terms:}

If the perturbative superpotential contains a term $h\Phi_1\Phi_2\Phi_3$,
then a trilinear scalar term appears in the effective Lagrangian
after SUSY breaking. It has the form
\begin{eqnarray}
{\cal L}_{tril} & = & -m_{3/2}\ A\ \hat h\phi_1\phi_2\phi_3
+ {\mathrm{h.c.}}
\nonumber \\
& = & -m_{3/2} \left\{
\left(1-Y\frac{\bar{W}_{\bar S}}{\bar{W}}\right)
 + \frac{Y}{3Y+\frac{\delta^{GS}}{4\pi^{2}}} \right.
\nonumber \\
& \times & \left(3+
\frac{\delta^{GS}}{4\pi^{2}}\frac{\bar W_{\bar S}}{\bar W}-(T+\bar{T})
\frac{\bar W_{\bar T}}{\bar W} \right) (3+\sum_{\alpha=1}^3 n_{\alpha})
\nonumber \\
& - & \left.  \frac{Y(T+\bar{T})}{3Y+\frac{\delta^{GS}}{4\pi^{2}}}
\left(3+\frac{\delta^{GS}}{4\pi^{2}}\frac{\bar W_{\bar S}}{\bar W}-
(T+\bar{T})\frac{\bar W_{\bar T}}{\bar W} \right)
\frac{h_{T}}{h}  \right\} \hat h \phi_{1}\phi_{2}\phi_{3} + {\mathrm {h.c.}}
\;\;,
\label{Ltri}
\end{eqnarray}
where $h_T\equiv \partial h/\partial T$,
$\phi_\alpha$  are the (properly normalized) scalar
components of the respective superfields, $n_\alpha$
are the corresponding modular weights, and  $\hat h =
e^{K/2}\prod_{\alpha=1}^3 (K_{\alpha \bar \alpha})^{-1/2}h$ is the
effective Yukawa coupling between the physical fields.
Notice that for the untwisted case
[$n_\alpha=-1$, $h\neq h(T)$] the previous expression is drastically
simplified.

\vspace{0.3cm}
\noindent {\em Bilinear scalar terms:}

It is not clear by now where a bilinear term
of the form $\mu H_1H_2$ in $W_{{\mathrm {obs}}}$ has its origin,
although it is well known that
this term is necessary in order to break $SU(2)\times U(1)_Y$
successfully\footnote{A possible origin for $\mu$ has been
proposed in ref.\cite{kim}. A trilinear term of the form $NH_1H_2$,
$N$ being an $SU(3)\times SU(2)\times U(1)_Y$ singlet with non--vanishing
expectation value could also do the job \cite{nil2}.}. This is the
so--called $\mu$ problem. Assuming that this term is actually present,
the corresponding bilinear term in the  scalar Lagrangian turns
out to be
\begin{eqnarray}
{\cal L}_{bil} & = & -m_{3/2}B \hat\mu H_1H_2 + \mathrm{h.c.} \nonumber \\
               & = & -m_{3/2}Y
 \left\{-\frac{\bar{W}_{S}}{\bar{W}}+
\frac{3+n_{1}+n_{2}}{3Y+\frac{\delta^{GS}}{4\pi^{2}}} \left(3+
\frac{\bar{W}_{\bar S}}{\bar{W}} \frac{\delta^{GS}}{4
\pi^{2}} \right. \right.
                -  \left. \left.
\frac{\bar{W}_{\bar T}}{\bar{W}}
(T+\bar{T})\right) \right.
\nonumber \\
               & - & \left. \frac{(T+\bar{T})}
{3Y+\frac{\delta^{GS}}{4 \pi^{2}}} \left[ 3 +
\frac{\delta^{GS}}{4 \pi^{2}} \frac{\bar{W}_{\bar S}}{\bar{W}}
-\frac{\bar{W}_{\bar T}}{\bar{W}}
(T+\bar{T}) \right] \frac{\mu_{T}}{\mu}\right\}
\hat\mu H_{1}H_{2}\; +\; \mathrm{h.c.}
\label{Lbil}
\end{eqnarray}
with a notation similar to that of eq.(12). Again,
$\hat \mu =
e^{K/2}\prod_{\alpha=1}^2 (K_{H_\alpha \bar H_\alpha})^{-1/2}
\mu$ is the
effective parameter giving, for example, the coupling between
the two higgsinos. For simplicity in the notation, we will drop
the hat in $\hat\mu$ as well
as in $\hat h$ from here on. Because of the above
mentioned ignorance about the origin of $\mu$, we prefer in the
following to consider $B$ as an unknown parameter.

A preliminary conclusion at this stage is that some of the
common assumptions of the minimal supersymmetric standard model,
in particular universality for all the gaugino masses, scalar masses
and trilinear terms, do not hold in general (this
was already noted in ref.\cite{ib1}). Nevertheless, it is
clear that eqs.(\ref{M12}--13) are not completely significant
unless we have a way to evaluate $m_{3/2}$, $\langle W\rangle$,
$\langle W_S\rangle$, etc.
This can only be done in the framework of a SUSY breaking scenario,
by minimizing the corresponding scalar potential. We turn to this
point in the next section.

\section{Realization of SUSY breaking by gaugino condensation}

As was mentioned above, the only mechanism so far analysed,
capable of generating a hierarchical SUSY breakdown in superstring
theories, is gaugino condensation in the hidden sector \cite{nil3}.
An extensive study of the properties of this mechanism can be
found in ref.\cite{dec2}. Let us briefly summarize here the main
characteristics. For a hidden sector gauge group $G=\prod_bG_b$,
gaugino condensation induces a non--perturbative superpotential
\begin{eqnarray}
W^{np}=\sum_b d_b\ \frac{e^{-3k_bS/2\beta_b}}{[\eta(T)]^{6-
(3k_b\delta^{GS}/4\pi^2\beta_b)}}
\;\;,
\label{Wcond}
\end{eqnarray}
where $\beta_b$ are the corresponding beta functions and $d_b$ are
constants. Notice that $W^{np}_T=-2\eta^{-1}\frac{\partial \eta}{\partial T}
(3W^{np}+\frac{\delta^{GS}}{4\pi^2}W^{np}_S)$, which can be used to eliminate
$W_T$ in (\ref{M12}--13). In particular,
\begin{eqnarray}
M_a(Q) & = & 2\pi\alpha_{a}(Q) m_{3/2} \left\{ k^{a}Y^{2}
\left(-\frac{1}{Y}+\frac{\bar{W}_{\bar S}}{\bar W}\right)
- \frac{Y(T+\bar T)^2}{3Y+\frac{\delta^{GS}}{4 \pi^2}}
\frac{|\hat{G}_{2}(T)|^2}{2\pi^2} \right.
 \nonumber \\
 & \times  & \left. \left(
\sum_{i} \frac{b_i^{'a}}{16 \pi^2}-
k^a\frac{\delta^{GS}}{8 \pi^2}\right)
\left(
3+\frac{\delta^{GS}}{4 \pi^2}\frac{\bar{W}_{\bar S}}{\bar W}
\right)\right\}
\label{M122}
\end{eqnarray}
\begin{eqnarray}
m_{\phi_\alpha}^2 =  m_{3/2}^{2}\left[
1+n_{\alpha}\left(\frac{T+\bar T}{2\pi}\right)^2\ |\hat G_2(T)|^2
\left|\frac{3+\frac{\delta^{GS}}{4\pi^2}\frac{W_S}{W} }
{3+\frac{\delta^{GS}}{4\pi^2}\frac{1}{Y}}\right|^2  \right]
\;\;,
\end{eqnarray}
There is a large class of scenarios for which SUSY is spontaneously
broken, yielding reasonable values of $m_{3/2}$ and $\langle Y \rangle$
without the need of fine tuning \cite{dec2}. A common ingredient of these
models is the existence of more than one condensing group
in the hidden sector \cite{kras,dix1,cas1}. The values of $m_{3/2}$ and
$\langle Y \rangle$ turn out to depend almost exclusively on
which the gauge group and matter content of the hidden sector are.
Practically, any values of $m_{3/2}$ and $\langle Y \rangle$ are available by
appropriately choosing the hidden sector. Consequently, we can consider
$m_{3/2}$ and $Y$ as free parameters since no dynamical mechanism
is already known to select a particular string vacuum. On the
other hand, both $m_{3/2}$ and $\langle Y \rangle$ are almost independent
of the value of $\delta^{GS}$ \cite{dec2}. Likewise, the value of
$\langle T \rangle$ turns out to be $\sim 1.23$ in all cases
\cite{font,dec2}. This stability does not hold, however, for the other
quantities ($\langle W \rangle$, $\langle W_S \rangle$, etc.) appearing in
eqs.(\ref{M12}--13) and (\ref{M122},\ref{mQ2}). For example,
as was pointed out in ref.\cite{font}, the combination
$YW_S-W$ (which appears in several places in
the previous equations) is vanishing in the minimum of the
scalar potential for $\delta^{GS}=0$, but this is no longer true for
$\delta^{GS}\neq 0$. In this case, however, the scalar potential
is much more involved, so a numerical analysis is in general
necessary.

Fortunately, the results admit quite simple and useful
parametrizations describing them very well (within $1\%$ of accuracy).
Next, we give these parametrizations for the common $k^a=1$ case.
In particular, for the scalar masses
\begin{eqnarray}
m_{\phi_\alpha}^2= m_{3/2}^2\left[1+n_\alpha(0.078-1.3\times 10^{-4}
\ \delta^{GS})\right]
\;\;.
\label{mapr}
\end{eqnarray}
It is really remarkable here the lack of dependence of (\ref{mapr})
on the specific type of hidden sector considered. Actually, the influence
of the hidden sector is completely summarized in the value of $m_{3/2}$.
Moreover, the influence of the actual values of the modular weights
$n_{\alpha}$ for the observable particles is quite small.

On the contrary, the absolute value of $A$, i.e. the parameter defining
the trilinear scalar terms in (12), has a stronger
dependence on the modular weights of the relevant fields. In particular,
if the three fields under consideration are untwisted, i.e.
$n_1=n_2=n_3=-1$, $A$ turns out to be very small in all the cases
\begin{eqnarray}
A^{untw}\stackrel{<}{{}_\sim} 10^{-3}<<1
\label{Aapr}
\end{eqnarray}
(if $\delta^{GS}=0$, $A^{untw}=0$ exactly).
For twisted fields a good parametrization is
\begin{eqnarray}
A = A^{untw} + (0.28-2.3\times10^{-4} \delta^{GS})\left(3+\sum_{\alpha=1}^3
n_{\alpha}\right) + (0.69-6.9\times10^{-4} \delta^{GS}) \frac{h_{T}}{h}
\;\;,
\label{Atot}
\end{eqnarray}
where the precise value of $h_{T}/h$ depends on the specific
Yukawa coupling considered (see ref.\cite{ham}). In general $h_{T}/h$ is
negative and $|h_{T}/h|\stackrel{<}{{}_\sim}O(1)$.

Concerning gaugino masses, it is worth noticing that
for $\delta^{GS}= 0$ the first term of eq.(\ref{M122})
(proportional to the $S$ F--term) vanishes in the minimum
of the potential \cite{font}. Then the resulting gaugino mass is very tiny
(e.g. for all the observable matter in the untwisted sector
the gluino mass is $M_3=m_{3/2}/445$).
This is not surprising since it is generated thanks to the one--loop
threshold correction of eq.(\ref{fw}). Fortunately, this fact does
not hold for $\delta^{GS}\neq 0$.
The values of the gaugino masses present a certain
dependence on the type of orbifold considered. Consider first the case of
the $Z_3$ and $Z_7$ orbifolds. For these schemes the threshold contributions
to the $f$ function [see eq.(\ref{fw})] are known to vanish \cite{kapl} and the
following equality holds for all the gauge group factors
\begin{eqnarray}
\sum_{i} b_i^{'a}\ -\ 2k^a\delta^{GS}=0
\;\;.
\label{cancel}
\end{eqnarray}
As a consequence, there is a cancellation of the second term of
eq.(\ref{M122}). The value of the first term at the minimum of
the potential is easily parametrizable in the form
\begin{eqnarray}
M_a(Q) = -\alpha_{a}(Q)\ m_{3/2}\ (0.0120\ \delta^{GS}+0.0029)
\;\;.
\label{Mapr}
\end{eqnarray}
Hence, for the $Z_3$ and $Z_7$ orbifolds the value of $M_a$
is completely given in terms of $m_{3/2}$ and $\delta^{GS}$. For
the rest of the $Z_N$ orbifolds things are different since eq.(\ref{cancel})
is no longer true. Then, an equation similar to eq.(\ref{Mapr})
can be written in terms of $m_{3/2}$, $\delta^{GS}$ and
$\sum_i b_{i}^{'a}$. Let us note, however, that for all the remaining
$Z_N$ orbifolds (except for the $Z_6$--II) the cancellation (\ref{cancel})
still takes place for two of the three compactified dimensions\footnote{
For the $Z_6$--II case the cancellation holds in one complex dimension.
On the other hand, the $Z_6$--II orbifold is one of the less interesting
orbifolds from the phenomenological point of view \cite{cas3}.}, i.e.
\begin{eqnarray}
b_i^{'a}\ -\ 2k^a\delta_i^{GS}=0\;\;,\;\;\;\;i=1,2
\;\;.
\label{cancel2}
\end{eqnarray}
Roughly speaking, our numerical results indicate that
parametrization (\ref{Mapr}) is still valid in these cases within
a $30\%$ error. As will be seen in the next section,
this is enough for our purposes. Finally, as explained above, we choose
to leave the value of $B$ [see eq.(13)] free, owing to our
ignorance of the origin of $\mu$. Equations (\ref{mapr},\ref{Atot},\ref{Mapr})
summarize the input of soft breaking terms to
be studied from the phenomenological point of view. Before entering
in this analysis, let us comment two features relative to
eqs.(\ref{mapr},\ref{Mapr}). First, an interesting $m_{3/2}$--independent
relation between $M_a$ and $m_{\phi_\alpha}$ can be obtained
\begin{eqnarray}
\frac{M_a^2(M_Z)}{m_{\phi_\alpha}^2(M_{Str})} = \frac{\alpha_{a}^2(M_Z)\
(0.0120\ \delta^{GS}+0.0029)^2}
{1+n_\alpha(0.078-1.3\times 10^{-4}
\ \delta^{GS})}
\;\;.
\label{Mm}
\end{eqnarray}
Of course, in order to get the physical scalar masses, an RG
running until the electroweak scale has to be performed. However,
it is already clear that gaugino masses tend to be much
smaller than scalar masses, since typically
$\delta^{GS}\stackrel{<}{{}_\sim} 50$ \cite{der}. Second, as was mentioned
in section 2, $M_a$ and $m_{\phi_\alpha}^2$, as well as the trilinear
terms, clearly show a certain lack of universality, unlike the usual
assumptions of the minimal supersymmetric standard model.
%
%
%
%
%

\section{Phenomenological viability of the soft breaking terms}
\subsection{Experimental constraints}
As was mentioned in the introduction, there are two types of
constraints on soft breaking terms: observational bounds and naturalness
bounds. The first ones mainly come from direct production of
supersymmetric particles in accelerators. This gives the following
lower bounds on supersymmetric particle masses, as reported by the
Particle Data Group \cite{par}
\begin{eqnarray}
M_3 &>&79\ \mathrm{GeV}\;\;,\;\;\; M_{\chi^{\pm}}{>}45\ \mathrm{GeV}
\nonumber \\
m_{\tilde q}&>&74\ \mathrm{GeV}\;\;,\;\;\; m_{\tilde l}
\stackrel{>}{{}_\sim} 45\ \mathrm{GeV}
\;\;,
\label{mexp}
\end{eqnarray}
where $M_3$ is the gluino mass, $\chi^{\pm}$ is the lightest
chargino, and $\tilde q, \tilde l$ collectively denote squarks and sleptons
respectively\footnote{Less conservative bounds have been reported
elsewhere \cite{des}. The corresponding modification of our final results
is completely straightforward.}. A more stringent lower bound for the
gluino and squark masses ($106\ $GeV) arises if $m_{\tilde q}=M_3$, which
certainly is not the case here. The chargino bound is subjected
to the condition $m_{\chi^o}<28\ $GeV with $\chi^o$ the lightest
neutralino, thus this bound does not necessarily apply.
The above experimental bounds are to be confronted with the
theoretical values from the SUSY breaking scenario we are analyzing.
In particular, the gluino mass is simply given by
eq.(\ref{Mapr}) evaluated at the $M_Z$ scale
\begin{eqnarray}
M_3 \simeq \alpha_{3}(M_Z) m_{3/2}\ (0.0120\ \delta^{GS}+0.0029)
\;\;.
\label{M3apr}
\end{eqnarray}
On the other hand, charginos (i.e. winos $\tilde W^{\pm}$ and higgsinos
$\tilde H^{\pm}$) mix via a $2\times 2$ mass matrix whose lowest
eigenvalue is given by
\begin{eqnarray}
M_{\chi^{\pm}}&=&\frac{1}{2}\left[M_2^2+\mu_R^2+2M_W^2 \right.
\nonumber \\
& - & \left.\sqrt{(M_2^2-\mu_R^2)^2 + 4M_W^4\cos^22\beta +
4M_W^2(M_2^2+\mu_R^2+2M_2\mu_R\sin2\beta)}\ \right]
\label{Mchar}
\end{eqnarray}
where ${\mathrm{tg}} \beta=\langle H_2\rangle/\langle H_1\rangle$,
$\mu_R$ is the renormalized value of $\mu$,
and $M_2$ is given by eq.(\ref{Mapr}). $M_2$
and $\mu_R$ are evaluated at the $M_Z$ scale.
In all these equations the current central experimental values
of $\alpha_a(M_Z)$ can be used
\begin{eqnarray}
\alpha_1(M_Z)^{exp}=0.01693,\;\;\alpha_2(M_Z)^{exp}=0.03395,
\;\;\alpha_3(M_Z)^{exp}=0.125
\;\;.
\label{alfexp}
\end{eqnarray}
%
Squark and slepton masses in our scenario are given (at $M_{Str}$)
by eq.(\ref{mapr}) where, of course, a standard RG running is necessary
in order to obtain the physical masses. Furthermore, for the left
and right top-squarks there is a non--trivial mixing in the mass
matrix which has to be conveniently taken into account. Nevertheless,
as was mentioned in the previous section, gaugino masses
in our framework tend to be clearly smaller than scalar masses, so the
relevant experimental (lower) bounds for us are those corresponding to
$M_3$ and $M_{\chi^{\pm}}$. In particular, comparing the theoretical gluino
mass (\ref{M3apr}) with the experimental limit (\ref{mexp}) we obtain
the following bound on $m_{3/2}$, $m_{\tilde q}$, $m_{\tilde l}$,
\begin{eqnarray}
m_{3/2} & \stackrel{>}{{}_\sim} & \frac{79\ \mathrm{GeV}}
{\alpha_{3}(M_Z)\ (0.0120\ \delta^{GS}+0.0029)}
\nonumber \\
m_{\tilde{q}}(M_{Str}), m_{\tilde{l}}(M_{Str})
& \stackrel{>}{{}_\sim} & (79\ \mathrm{GeV})
\frac{\sqrt{1+n_{\alpha}(0.078-1.3\times10^{-4}
\delta^{GS})}}{\alpha_{3}(M_{Z})(0.0120 \delta^{GS}+0.0029)}
\;\;,
\label{bound1}
\end{eqnarray}
where we have used eq.(\ref{mapr}). Similar bounds, but
involving also $\mu$ and $\sin 2\beta$, are obtained from the lightest
chargino mass (\ref{Mchar}).

Before going into the naturalness constraints, let us remark that
there exits an upper bound on the value of $|A_t|$ (i.e. the parameter of
the trilinear scalar term coming from the top Yukawa coupling) in order
to prevent the scalar potential of developing colour and electric charge
breaking minima. Assuming universality for the SUSY breaking
scalar masses ($m_\phi=m$), this bound reads $A_t^2\leq 3(3+\mu^2/m^2)$
at the GUT scale \cite{des2}. Hence, for $\mu^2<<m^2$ (which will be
the case here) we can simply use $|A_t|\leq 3$.
This constraint is trivially satisfied if the top Yukawa coupling
is of the untwisted type, see eq.(\ref{Aapr}). In general, $|A_t|$
is given by eq.(\ref{Atot}) and is small enough to avoid the
existence of these troublesome vacua.

\subsection{Naturalness constraints}

The naturalness constraints come from requiring the absence of fine tuning
in the electroweak breaking mechanism. Let us review very briefly
the previous work on this issue.
The standard criterion \cite{bar} to avoid
the fine tuning is to impose
\begin{eqnarray}
\left|\frac{a_i}{M_Z^2}\frac{\partial M_Z^2(a_i,h_t)}
{\partial a_i}\right|<\Delta
\;\;,
\label{natur}
\end{eqnarray}
where $a_i$ are the theoretical parameters defining the SUSY breaking,
$h_t$ is the top Yukawa coupling, which is taken as an independent parameter,
and $\Delta$ measures the allowed degree of fine tuning.
In ref.\cite{bar} the soft breaking parameters $M, m, \mu, A, B $ were
taken as the $a_i$ parameters of eq.(\ref{natur}), assuming universality
of the SUSY breaking gaugino and scalar masses at the unification scale
(i.e. $m_{\phi_\alpha}\equiv m$, $M_a\equiv M$ $\forall a,\alpha$).
The relevant expression for $M_Z^2$ is
\begin{eqnarray}
M_{Z}^2
=\frac{2(\mu_1^2-\mu_2^2\mathrm{tg}^2 \beta)}{\mathrm{tg}^2\beta-1}
\;\;,
\label{MZ}
\end{eqnarray}
where the $\beta$ angle, defined in eq.(\ref{Mchar}),
is given by $\sin 2\beta=2\mu_3^2/(\mu_1^2+\mu_2^2)$.
This comes from minimizing the neutral part of the Higgs potential
\begin{eqnarray}
V(H_1,H_2)=\frac{1}{8}(g^2+g'^2)\left(|H_1|^2-|H_2|^2\right)^2
+ \mu_1^2|H_1|^2 + \mu_2^2|H_2|^2 -\mu_3^2(H_1H_2+\mathrm{h.c.})
\;\;,
\label{Vhiggs}
\end{eqnarray}
where $\mu_1$, $\mu_2$, $\mu_3$ are related to the soft breaking parameters
via well--known expressions \cite{ib2} coming from integrating the
corresponding RG equations. More precisely,
\begin{eqnarray}
\mu_1^2 & = & m^2 + \mu^2l(t) + M^2g(t)
\nonumber \\
\mu_2^2 & = & m^2(h(t)-k(t)A^2) + \mu^2l(t) + M^2e(t) + AmMf(t)
\nonumber \\
\mu_3^2 & = & -\mu mBq(t) + \mu Mr(t) + Am\mu s(t)
\;\;,
\label{mus1}
\end{eqnarray}
where $l(t)$, $g(t)$, etc. are RG quantities whose explicit
dependence on the RG parameter $t=2\log(M_{GUT}/M_Z)$, $\alpha_a$ and
$h_t$ can be found in ref.\cite{ib2}, where the minimal particle content
and the usual approximation $h_b,h_\tau<<h_t$ were used.
Using eqs. (\ref{MZ}) and (\ref{mus1})
the naturalness conditions (\ref{natur}) for each soft breaking
parameter are easily obtained \cite{bar}. A further simplification consists
of working around the approximate zeros of eq.(\ref{MZ}), i.e.
\begin{eqnarray}
\mu_1^2\  \mu_2^2 \simeq\  \mu_3^4
\;\;.
\label{zeros}
\end{eqnarray}
This is equivalent to requiring the right value for $M_Z$ in (\ref{MZ}),
neglecting $\frac{M_Z^2}{M^2}$, $\frac{M_Z^2}{m^2}$,
$\frac{M_Z^2}{\mu^2}$ terms. Then the left hand side of
eq.(\ref{natur}) is simplified since
\begin{eqnarray}
\frac{a_i}{M_Z^2}\frac{\partial M_Z^2(a_i,h_t)}
{\partial a_i}=
\frac{2(\partial/\partial a_i)(\mu_1^2\mu_2^2-\mu_3^4)}
{M_Z^2(\mu_1^2-\mu_2^2)}
\;\;.
\label{natur2}
\end{eqnarray}

The scenario we are analysing is similar to the previous one,
but presents some peculiarities. First, the RG running has to
be made from $M_{Str}$ instead of
$M_{GUT}$. Likewise, neither the soft breaking parameters $m_{\phi_\alpha}^2$,
$M_a$, nor the gauge couplings $\alpha_a$ are "unified" at $M_{Str}$.
More precisely, $m_{\phi_\alpha}^2$ and $M_a$ are given by
eqs.(\ref{mapr},\ref{Mapr}), whereas $\alpha_a$ at $M_{Str}$ should
be obtained by inserting $\alpha_a^{exp}(M_Z)$ in the corresponding
RG equations, where we have taken the minimal particle
content too. All this modifies
the previous eqs.(\ref{mus1}) in a straightforward
way\footnote{The quantitative effect of these modifications, however, is
very small.}. Second, in our framework $M_a$, $m_{\phi_\alpha}^2$
and $A$ are not independent parameters since they are related
to $m_{3/2}$ and $\delta^{GS}$ through eqs.(\ref{mapr},\ref{Atot},\ref{Mapr}).
In consequence, eqs.(\ref{mus1}) become
\begin{eqnarray}
\mu_1^2 & = & C_1(\delta^{GS}) m_{3/2}^2 + D \mu^2
\nonumber \\
\mu_2^2 & = & C_2(\delta^{GS}) m_{3/2}^2 + D \mu^2
\nonumber \\
\mu_3^2 & = & \mu m_{3/2}\left[ C_3(\delta^{GS})+C_4B \right]
\;\;,
\label{mus2}
\end{eqnarray}
where the explicit expressions of $\ C_i,D\ $ can readily be obtained
from (\ref{mus1}) and (\ref{mapr},\ref{Atot},\ref{Mapr}) in the
above explained manner. For simplicity we have chosen here $H_1$
and $H_2$ as untwisted fields, thus $n_{H_1}=n_{H_2}=-1$ in
eq.(\ref{mapr}). Recall however that the actual dependence of
$m_{\phi_\alpha}^2$ on $n_\alpha$ is very small. Moreover, condition
(\ref{zeros}) (which in this case amounts to neglect
$\frac{M_Z^2}{m_{3/2}^2}$, $\frac{M_Z^2}{\mu^2}$ terms) reads
\begin{eqnarray}
\left(\frac{\mu}{m_{3/2}}\right)^2 & \simeq & - \frac{1}{2D^2}
\left[ D (C_1+C_2)-(C_3+C_4B)^2 \right.
\nonumber \\
 & \pm &
\left. \sqrt{(D(C_1+C_2)-(C_3+C_4B)^2)^2-4D^2C_1C_2}\right]\equiv f(B)
\;\;.
\label{zeros2}
\end{eqnarray}
Clearly, the role of the independent $a_i$ parameters in (\ref{natur})
has to be played here by $m_{3/2}$, $\mu$ and $B$. Consequently, our
naturalness conditions are
\begin{eqnarray}
m_{3/2} & : & \;\;\;\;\;\;
m_{3/2}^2\ <\ \frac{M_Z^2}{4}\left|\frac{C_1-C_2}
{2C_1C_2+f(B)(D(C_1+C_2)-(C_3+C_4B)^2)}\right|\Delta
\label{mnat} \\
\mu & : & \;\;\;\;\;\;
\mu^2\ <\ \frac{M_Z^2}{4}\left|\frac{C_1-C_2}
{2D^2f(B)+D(C_1+C_2)-(C_3+C_4B)^2}\right|\Delta
\label{munat} \\
B & : & \;\;\;\;\;\;
|B(C_3+C_4B)\mu^2|\ <\ \frac{M_Z^2}{4}\left|\frac{C_1-C_2}{C_4}
\right|\Delta
\;\;.
\label{Bnat}
\end{eqnarray}
In writing(\ref{mnat}) and (\ref{munat}) $\mu^2$ and $m_{3/2}^2$
have respectively been eliminated through (\ref{zeros2}).
Of course, the level of stringency of conditions
(\ref{mnat}--\ref{Bnat}) depends on the value of $\Delta$, which is,
to some extent, arbitrary. More precisely, the bounds
on $m_{3/2}$ and $\mu$ (and hence those on $M_a$ and
$m_{\tilde q}$) scale as $\sqrt{\Delta}$.
Notice also that if we had taken $M_Z$ instead
of $M_Z^2$ in eq.(\ref{natur}), then the
bounds would have been less restrictive by a factor of $\sqrt{2}$.
Therefore, we have adopted a rather conservative point of view,
taking $\Delta=50$.
Following ref.\cite{bar} the quantitative results are conveniently
given in terms of the quantity $M_t$, defined as
\begin{eqnarray}
M_t=\frac{\sqrt{2}}{g_2}M_W h_t
\;\;,
\label{Mt}
\end{eqnarray}
so that the physical top mass is given by
\begin{eqnarray}
m_t=M_t \sin \beta
\;\;.
\label{mtMt}
\end{eqnarray}

The positivity
of the right hand side of eq.(\ref{zeros2}), selects a definite
sign of the square root and, together with the naturalness
conditions (\ref{mnat}--\ref{Bnat}), imposes a lower bound on $M_t$,
say $M_t^{min}$ \footnote{This kind of limit did not appear
in ref.\cite{bar}, since the value of $B$ was always taken such that
eq.(\ref{zeros}) were fulfilled, without taking into account the
naturalness bounds for it.}.
This corresponds to the value of $M_t^{min}$ for
which the quantity $C_{2}$ is vanishing, which depends also on
the value of $A$. This is shown in fig.1: maintaining $|A| \leq
3$, $M_{t}^{min}$ is forced to be in the region
\begin{eqnarray}
80\ \mathrm{GeV} \leq M_{t}^{min} \leq 158\ \mathrm{GeV}
\label{Mtmin}
\end{eqnarray}
where the upper limit corresponds to the untwisted case.
The naturalness conditions
(\ref{mnat}--\ref{Bnat}) for a typical case ($\delta^{GS}=45$)
are illustrated in fig.2. More precisely, in
fig.2a the upper bounds
on $m_{3/2}$ and $\mu$, i.e. eqs.(\ref{mnat},\ref{munat}), have
been represented as functions of $M_t-M_t^{min}$. [The thickness of
the curves simply counts the
variation of $B$ within its naturalness limits, given by (\ref{Bnat})].
This is translated into upper bounds on $M_3$, $M_{\chi^\pm}$
through eqs.(\ref{M3apr},\ref{Mchar}), which is illustrated
in fig.2b. We have also represented the lower experimental
bounds on $M_3$, $M_{\chi^\pm}$ (dashed lines), so that the complete
allowed range of variation for $M_3$, $M_{\chi^\pm}$ and $M_t$
can be easily observed. Analogously, upper bounds on
squark and slepton masses are obtained through
eq.(\ref{mapr}). Note that this equation gives the masses
at $M_{Str}$, so a standard RG running
has been performed. The corresponding upper bounds are represented
in fig.2c. Just for simplicity we have considered here
the case of untwisted fields [i.e. $n_\alpha=-1$ in
eq.(\ref{mapr})], but the extrapolation to the general
case is straightforward. As was mentioned above, it is clear
that the experimental lower bounds on squark and slepton
masses do not play any role since, once $M_3$, $M_{\chi^\pm}$
are compelled to respect their experimental lower bounds, the former
are automatically fulfilled thanks to eq.(28).
In consequence, the relevant lower
bound for $m_{\tilde{q}}$, $m_{\tilde{l}}$ is the one coming from
eq.(28), which has been included in the figure
(dashed line). Note
that this bound arises from the experimental gluino limit
[see eq.(\ref{mexp})]. The one corresponding to the chargino
(which has not been represented) is a bit more stringent,
although it is subjected to the
above mentioned condition on the lightest neutralino mass.
Notice also that indeed $\mu^2<<m_\phi^2$, as is required for the
approximation $|A|\leq 3$.
Let us summarize the allowed ranges for the relevant masses
in the $\delta^{GS}=45$ case. Excluding values of
$M_{t}$ very close to the peak in fig.2 (which would amount to a
new fine tuning by itself\footnote{More precisely, we are allowing
for a fine tuning of $0.5\ $GeV in the value of $M_t$.}) these are
\begin{eqnarray}
1100\ \mathrm{GeV}\leq m_{3/2}\leq 4200\ \mathrm{GeV} \;& , & \;\;
350\ \mathrm{GeV}\leq \mu
\leq 450\ \mathrm{GeV}
\nonumber \\
79\ \mathrm{GeV}\leq M_3\leq 285\ \mathrm{GeV} \;& , & \;\; 22 \
\mathrm{GeV}\leq
M_{\chi^{\pm}}\leq 80\ \mathrm{GeV}
\nonumber \\
-7\ \mathrm{GeV}\leq M_{t}^{min}-m_t\leq 1\ \mathrm{GeV}\;& , & \;\;
1070\ \mathrm{GeV}\leq
m_{\tilde q}, m_{\tilde l}\leq 4080 \ \mathrm{GeV}
\;\;,
\label{cotas}
\end{eqnarray}
%
%
%
where $M_t^{min}$ is constrained by eq.(\ref{Mtmin}). All the
lower bounds in (\ref{cotas}) arise from the experimental gluino
limit in eq.(\ref{mexp}). The previous allowed ranges are narrowing
as $\delta^{GS}$ decreases, so that for $\delta^{GS}<25$ they
virtually disappear. Therefore, we conclude that $\delta^{GS}>25$ is
a necessary condition for respecting all the (experimental and
naturalness) bounds on supersymmetric masses. It is remarkable
in (\ref{cotas}) the smallness of the gaugino masses (quite close
to their experimental lower limits of eq.(\ref{mexp})) compared with
the squark and slepton masses. Furthermore, the top mass $m_t$
(see eqs.(\ref{cotas},\ref{Mtmin})) is notably consistent with
its present experimental bounds. Finally, let us recall that
the values in eq.(\ref{cotas}) have been obtained by using
eq.(\ref{Mapr}) for the gaugino masses, which is strictly valid
only for the $Z_3$, $Z_7$ orbifolds. Therefore, as was mentioned
in the text, the previous numbers can oscillate within a $30\%$
error in the general case. This does not substantially modify,
however, the main conclusions.

\section{Summary and conclusions}

We have calculated the form of the soft SUSY breaking terms arising from a
large class of string models, more precisely symmetric orbifold
constructions, and studied its phenomenological viability.
The form of the soft breaking terms depends on certain parameters,
such as the value of $\delta^{GS}$ (the contribution of the
Green--Schwarz mechanism to the gauge kinetic function) and the
modular weights of the relevant superfields. In the calculation
the cosmological constant has been assumed to vanish.
A first conclusion is that some of the
common assumptions of the minimal supersymmetric standard model,
in particular universality for all the gaugino masses, scalar masses
and trilinear terms, do not hold in general
(this was also noted in ref.\cite{ib1}).

In order to get quantitative results a specific SUSY breaking
mechanism has to be considered. In this sense the only mechanism
so far analysed, which is
capable of generating a hierarchical SUSY breakdown in string
constructions, is gaugino condensation in the hidden sector. Assuming
this scenario, we have given some explicit numerical formulae
for the various soft breaking terms. At this point it becomes
clear that squark and slepton masses tend to be much larger
than gaugino masses.

Finally, we have studied the phenomenological viability of the
soft breaking terms. There are two types of tests that they must pass.
First, they have to be consistent with the experimental (lower) bounds
on gaugino masses, squark masses, etc. Second, they should be small enough
not to spoil the SUSY solution to the gauge hierarchy problem, guaranteeing
a successful $SU(2)\times U(1)_Y\rightarrow U(1)_{em}$ breaking
without the need of fine tuning. We have found that the considered
scenarios can be perfectly consistent with both tests. This is
certainly non--trivial and in fact imposes interesting constraints on
measurable quantities (e.g. gaugino and scalar masses).
These constraints are summarized in the allowed ranges of
eq.(\ref{cotas}) for the $\delta^{GS}=45$ case. These are narrowing
as $\delta^{GS}$ decreases, so that for $\delta^{GS}<25$ they
virtually disappear. From eqs.(\ref{cotas},\ref{mexp}) it is clear
that gluino and chargino masses should not be much higher than
their present experimental lower bounds, while squark and slepton
masses must be much larger. All this can be
considered as an experimental signature of these stringy scenarios.
Also, the top mass is constrained to be
within a range [see eqs.(\ref{cotas},\ref{Mtmin})] remarkably
consistent with its present experimental bounds.

\vspace{2cm}
\noindent{\bf ACKNOWLEDGEMENTS}

We thank L.E. Ib\'{a}\~{n}ez for very useful comments and
suggestions. The work of
B.C. was supported by a Comunidad de Madrid grant.




\vspace{1.7cm}
\noindent{\bf FIGURE CAPTIONS}

\begin{description}

\item[Fig.1] $M_t^{min}$ (in GeV) versus $A$.

\item[Fig.2a] Upper bounds on $m_{3/2}$, $\mu$ (in GeV)
versus $M_t-M_t^{min}$ (in GeV). The thickness of
the curves counts the variation of $B$ within its naturalness limits.

\item[Fig.2b] Upper bounds on the gluino mass ($M_3$) and
the chargino mass ($M_{\chi^{\pm}}$) (in GeV)
versus $M_t-M_t^{min}$ (in GeV). The dashed lines represent
the corresponding lower experimental bounds. The one for
$M_{\chi^{\pm}}$ applies only if the lightest neutralino mass
is smaller than 28 GeV. The thickness of
the curves counts the variation of $B$ within its naturalness limits.

\item[Fig.2c] The same as in fig.2b but for the squark and slepton
masses ($m_{\tilde q}$, $m_{\tilde l}$). The dashed line represents
the lower bound coming from the experimental bound on
the gluino mass (see eq.(28)).

\end{description}

\end{document}